# SHIELDING EXPERIMENTS BY THE JASMIN COLLABORATION AT FERMILAB (II) - RADIOACTIVITY MEASUREMENT INDUCED BY SECONDARY PARTICLES FROM THE ANTI-PROTON PRODUCTION TARGET*[†]

Hiroshi YASHIMA[1#], Norihiro MATSUDA[2], Yoshimi KASUGAI[2], Hiroshi MATSUMURA[3], Hiroshi IWASE[3], Norikazu KINOSHITA[3,4], David BOEHNLEIN[5], Gary LAUTEN[5], Anthony LEVELING[5], Nikolai MOKHOV[5], Kamran VAZIRI[5], Koji OISHI[6], Hiroshi NAKASHIMA[2], and Yukio SAKAMOTO[2]

[1]Research Reactor Institute, Kyoto University, Kumatori, Sennan-gun, Osaka, 590-0494, Japan
[2]Japan Atomic Energy Agency, Tokai, Naka-gun, Ibaraki, 319-1195, Japan
[3]High Energy Accelerator Research Organization, Oho, Tsukuba, Ibaraki, 305-0801, Japan
[4]Tsukuba University, Ten'noudai, Tsukuba, Ibaraki, 305-8571, Japan
[5]Fermi National Accelerator Laboratory, Batavia, IL 60510, USA
[6]Shimizu Corporation, Etchujima, Koto-ku, Tokyo 135-8530, Japan

## Abstract

The JASMIN Collaboration has performed an experiment to conduct measurements of nuclear reaction rates around the anti-proton production (Pbar) target at the Fermi National Accelerator Laboratory (FNAL). At the Pbar target station, the target, consisting an Inconel 600 cylinder, was irradiated by a 120 GeV/c proton beam from the FNAL Main Injector. The beam intensity was $3.6 \times 10^{12}$ protons per second. Samples of Al, Nb, Cu, and Au were placed near the target to investigate the spatial and energy distribution of secondary particles emitted from it. After irradiation, the induced activities of the samples were measured by studying their gamma ray spectra using HPGe detectors. The production rates of 30 nuclides induced in Al, Nb, Cu, Au samples were obtained. These rates increase for samples placed in a forward (small angle) position relative to the target. The angular dependence of these reaction rates becomes larger for increasing threshold energy. These experimental results are compared with Monte Carlo calculations. The calculated results generally agree with the experimental results to within a factor of 2 to 3.

---

[*]Work supported by grand-aid of ministry of education (KAKENHI 19360432) in Japan and Fermi Research Alliance, LLC under contract No. DE-AC02-07CH11359 with the U.S. Department of Energy.
[†]Presented paper at International Conference on Nuclear Data for Science and Technology April 26-30, 2010, Jeju Island, Korea.
[#]yashima@rri.kyoto-u.ac.jp

# Shielding experiments by the JASMIN collaboration at Fermilab (II) - Radioactivity measurement induced by secondary particles from the anti-proton production target

Hiroshi YASHIMA[*], Norihiro MATSUDA[1], Yoshimi KASUGAI[1], Hiroshi MATSUMURA[2], Hiroshi IWASE[2], Norikazu KINOSHITA[2,3], David BOEHNLEIN[4], Gary LAUTEN[4], Anthony LEVELING[4], Nikolai MOKHOV[4], Kamran VAZIRI[4], Koji OISHI[5], Hiroshi NAKASHIMA[1], and Yukio SAKAMOTO[1]

Research Reactor Institute, Kyoto University, Kumatori, Sennan-gun, Osaka, 590-0494, Japan
[1]Japan Atomic Energy Agency, Tokai, Naka-gun, Ibaraki, 319-1195, Japan
[2]High Energy Accelerator Research Organization, Oho, Tsukuba, Ibaraki, 305-0801, Japan
[3]Tsukuba University, Ten'noudai, Tsukuba, Ibaraki, 305-8571, Japan
[4]Fermi National Accelerator Laboratory, Batavia, IL 60510, USA
[5]Shimizu Corporation, Etchujima, Koto-ku, Tokyo 135-8530, Japan
[*]Corresponding author. E-mail: yashima@rri.kyoto-u.ac.jp

The JASMIN Collaboration has performed an experiment to conduct measurements of nuclear reaction rates around the anti-proton production (Pbar) target at the Fermi National Accelerator Laboratory (FNAL). At the Pbar target station, the target, consisting an Inconel 600 cylinder, was irradiated by a 120 GeV/c proton beam from the FNAL Main Injector. The beam intensity was $3.6 \times 10^{12}$ protons per second. Samples of Al, Nb, Cu, and Au were placed near the target to investigate the spatial and energy distribution of secondary particles emitted from it. After irradiation, the induced activities of the samples were measured by studying their gamma ray spectra using HPGe scintillation counters. The production rates of 30 nuclides induced in Al, Nb, Cu, Au samples were obtained. These rates increase for samples placed in a forward (small angle) position relative to the target. The angular dependence of these reaction rates becomes larger for increasing threshold energy. These experimental results are compared with Monte Carlo calculations. The calculated results generally agree with the experimental results to within a factor of 2 to 3.

## 1. INTRODUCTION

The Japanese and American Study of Muon Interaction and Neutron detection (JASMIN) collaboration has been organized to study radiation effects associated with the high energy particle beams at the Fermi National Accelerator Laboratory (FNAL) [1]. The collaboration aims to study the behavior of secondary particles generated from beam losses in high-energy accelerators: particle fluxes and spectra, activities in air, water and materials, and radiation damage of components. Although considerable work has been done by similar experiments in the past [2-5], only a few have attempted such measurements with beam energies exceeding 1 GeV [6, 7]. In this energy region, most simulation codes implement a transition of reaction models from intermediate to high energies [8-11]. The collection of reaction data and comparison of the measured rates to simulations is the basic methodology for benchmarking studies to improve the reaction models.

The data discussed here were taken at the anti-proton (Pbar) production hall at FNAL. The Pbar target produces anti-protons for FNAL experiments in the Tevatron collider using a 65-kW proton beam of momentum 120 GeV/c. The spatial distribution of nuclear interactions and the neutron flux behind the shielding have been measured at this facility and an analysis comparing them with Monte Carlo calculations has been performed. Preliminary results are reported [12-15] and the studies are ongoing.

In this study, the spatial distribution of secondary particles generated in the Pbar target is inferred from the reaction rate in samples placed nearby. The experimental setup is described in Section 2. The results are discussed in Section 3, where they are compared with a calculation using the PHITS code [10, 11].

## 2. EXPERIMENT

**Fig. 1** shows a cross-sectional view of the Pbar target station. At this station, an anti-proton production target made of Inconel 600 is irradiated by 120 GeV protons. The average beam intensity is $3.6 \times 10^{12}$ protons/sec. Downstream of the target, a collection lens, collimator and a pulsed magnet focus, collimate, and extract the anti-protons produced in the target. The remaining protons and the secondary particles are absorbed by a beam dump downstream of the pulsed magnet. Iron and concrete shields, of thickness 188 cm and 122 cm respectively, are placed above the target and magnets. An air gap of 179 cm separates the iron and concrete shields. The distance between the target and the iron shields is 46 cm. The activation foils (Al, Nb, Cu, Au) were set around the anti-proton target to investigate the spatial and energy distribution of secondary particles emitted from the target. Three sets of foils, called TG-A, B, C, were placed at angles of 24.4°, 42.3° and 53.7° respectively, relative to the proton beam direction. **Fig. 2** shows the experimental set up around the target.

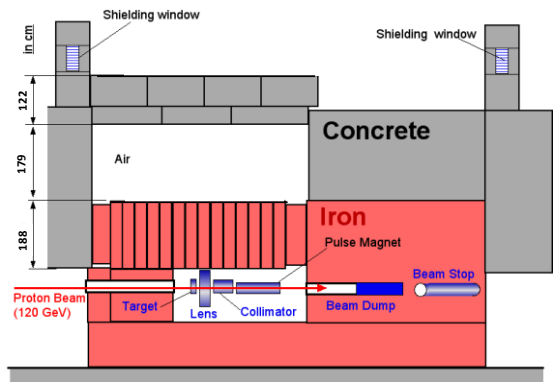

**Fig. 1.** The cross sectional view of the Pbar target station

After irradiation, the samples were taken to a counting lab equipped with Canberra high-purity germanium (HPGe) detectors. The HPGe counters are used to characterize the gamma-ray spectra and measure the decay curves. In this way, the nuclides induced in the foils are identified. Reaction rates for each nuclide are determined after being corrected for the peak efficiency of the HPGe detectors and the beam current fluctuation during the irradiation. One of the HPGe detectors was calibrated by Canberra. The peak efficiency of the calibrated HPGe detector was determined using Canberra's LabSOCS software [16]. For the non-calibrated detectors, the peak efficiency was determined from the ratios of their peak counting rates to that of the calibrated detector. The components of estimated errors in reaction rates were the counting statistics and detector efficiency.

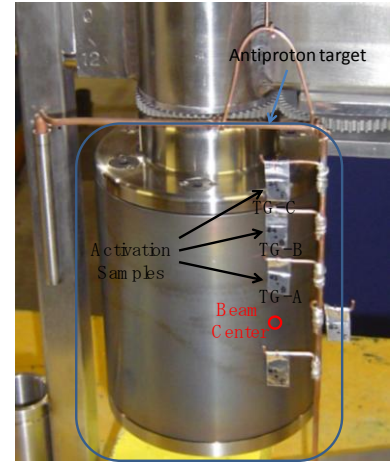

**Fig. 2.** The experimental set up showing the placement of activation foils near the antiproton production target.

## 3. RESULTS AND DISCUSSION

The reaction rates, expressed in product atoms per target atom per POT, of 30 nuclides were measured for the Al, Nb, Cu, Au samples. The activation products and their characteristics are listed in **Table 1** [17].

**Fig. 3** shows the angular dependence of the reaction rates. The vertical axis is the ratio of the reaction rate relative to that for TG-A. The horizontal axis is the threshold energy for neutron-induced reactions for the activation products shown in the legend. In this study, threshold energy for each spallation product is figured out by assuming that the rest of the target nucleus is emitted as a single nucleon.

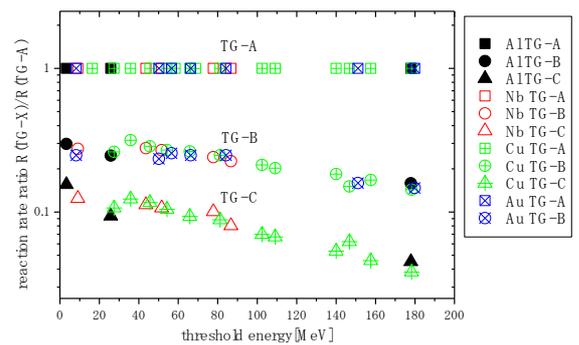

**Fig. 3.** The angular dependence of reaction rate of nuclides induced in samples

**Fig. 3** clearly shows the reaction rate decreasing at larger angles relative to the proton beam. This angular dependence becomes more pronounced as the threshold energy increases.

These experimental results are compared with predictions calculated with the PHITS code. **Fig. 4** shows the geometry for PHITS calculation. **Fig. 5** shows the calculated energy spectra of secondary neutrons, protons, $\pi^+$ and $\pi^-$ at the foil locations. The three curves show a decrease in the flux of secondary particles at larger angles. This angular dependence increases at higher energies. This calculation is consistent with the experimental results shown in **Fig. 3**.

**Fig. 6** shows the comparison of measured reaction rates of nuclides induced in Al, Nb, Cu and Au samples. The reaction rates of nuclides induced in samples around the target are expressed in units of reactions per nuclide per proton-on-target.

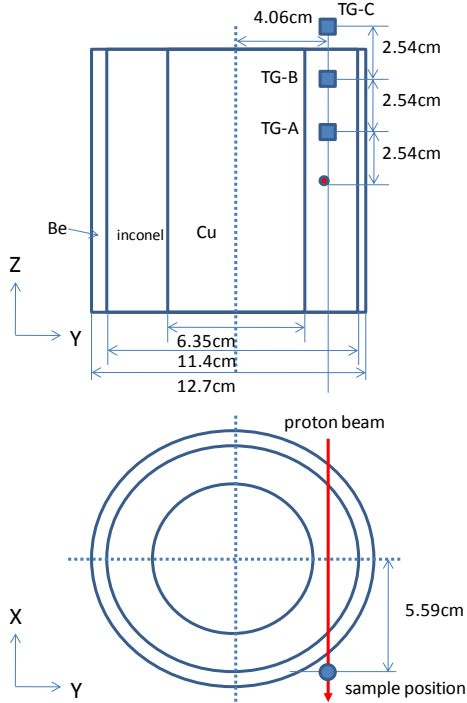

**Fig. 4.** The geometry for PHITS calculation.

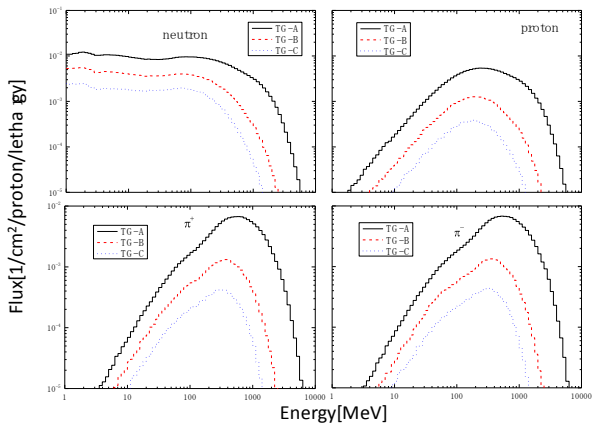

**Fig. 5.** The energy spectrum of neutrons, protons, $\pi^+$ and $\pi^-$ emitted from the anti-proton target calculated from PHITS.

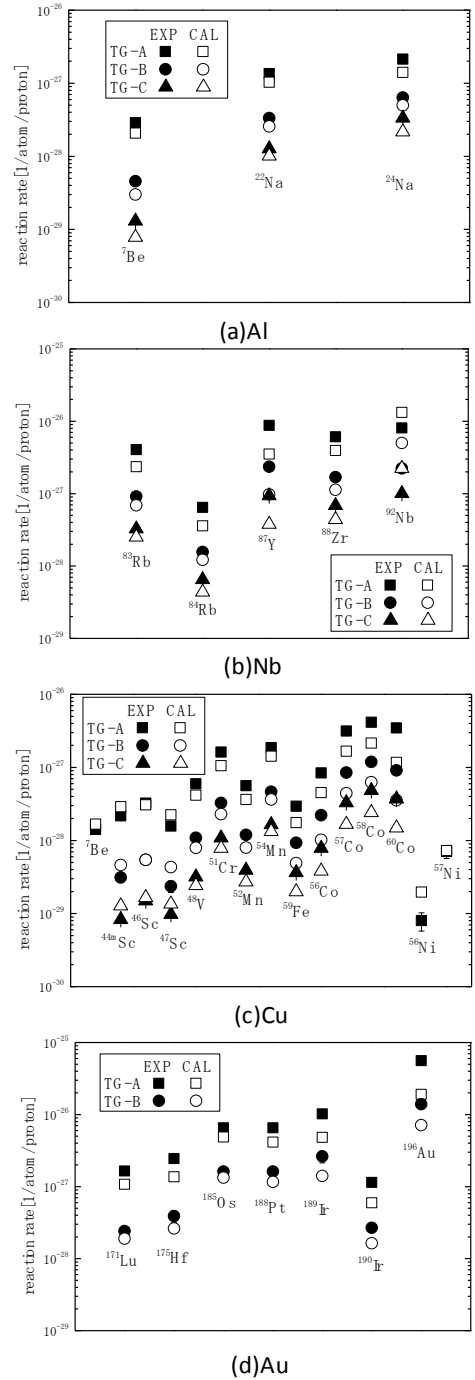

**Fig. 6.** The comparison of measured reaction rates of nuclides induced in Al, Nb, Cu and Au samples

In **Fig.6**, the calculated results generally underestimate the experimental results by a factor of 2 to 3, except for several products. For the case of $^{92}$Nb and $^{44}$Sc, the activity of the metastable nuclide was measured in the experiment, but the calculated results include both the metastable and ground-state nuclides.

**Table 1** The physical properties of radioactive nuclides [17]

| Sample | Product | Half-life | Gamma-ray Energy[keV] | Branching Ratio[%] |
|---|---|---|---|---|
| Al | $^{7}$Be | 53.29D | 477.595 | 10.5 |
| Al | $^{22}$Na | 2.6019Y | 1274.53 | 99.94 |
| Al | $^{24}$Na | 14.959H | 1368.633 | 100 |
| Nb | $^{83}$Rb | 86.2D | 520.389 | 47.7 |
| Nb | $^{84}$Rb | 32.77D | 881.61 | 69.0 |
| Nb | $^{87}$Y | 79.8H | 484.805 | 89.74 |
| Nb | $^{88}$Zr | 83.4D | 392.9 | 100 |
| Nb | $^{92m}$Nb | 10.15D | 520.389 | 47.7 |
| Cu | $^{7}$Be | 53.29D | 477.595 | 10.5 |
| Cu | $^{44m}$Sc | 58.6H | 271.13 | 86.8 |
| Cu | $^{46}$Sc | 83.79D | 889.277 | 99.98 |
| Cu | $^{47}$Sc | 3.345D | 159.381 | 67.9 |
| Cu | $^{48}$V | 15.9735D | 944.132 | 7.76 |
| Cu | $^{51}$Cr | 27.702D | 320.1 | 9.86 |
| Cu | $^{52}$Mn | 5.591D | 744.233 | 90.6 |
| Cu | $^{54}$Mn | 312.3D | 834.848 | 99.98 |
| Cu | $^{59}$Fe | 44.503D | 1099.251 | 56.5 |
| Cu | $^{56}$Co | 77.27D | 846.771 | 99.94 |
| Cu | $^{57}$Co | 271.79D | 122.0614 | 85.6 |
| Cu | $^{58}$Co | 70.82D | 810.775 | 99.45 |
| Cu | $^{60}$Co | 5.2714Y | 1332.501 | 99.98 |
| Cu | $^{56}$Ni | 5.9D | 749.95 | 49.5 |
| Cu | $^{57}$Ni | 35.6H | 1377.63 | 81.7 |
| Au | $^{171}$Lu | 8.24D | 739.78 | 47.8 |
| Au | $^{175}$Hf | 70D | 343.4 | 84.0 |
| Au | $^{185}$Os | 93.6D | 646.116 | 78.0 |
| Au | $^{188}$Pt | 10.2D | 195.05 | 18.6 |
| Au | $^{189}$Ir | 13.2D | 245.08 | 6.0 |
| Au | $^{190}$Ir | 11.78D | 605.14 | 39.9 |
| Au | $^{196}$Au | 6.183D | 335.68 | 86.9 |

## 4. CONCLUSION

The spatial distributions of activation rates from secondary particles near the FNAL antiproton target station have been measured. An angular dependence is found and the rates are higher in the forward direction. The angular dependence increases as the threshold energy for activation becomes higher. The calculated results by PHITS code agree with the experimental results to within a factor of 2 to 3. These experimental results will be useful as benchmark data to investigate the accuracy of various transport calculation codes.

## ACKNOWLEDGMENTS

This work is supported by grand-aid of ministry of education (KAKENHI 19360432) in Japan. Fermilab is a U.S. Department of Energy Laboratory operated under Contract DE-AC02-07CH11359 by the Fermi Research Alliance, LLC.